\documentclass[11pt]{article}

\usepackage{moriond,epsfig}

\begin{document}

\bibliographystyle{unsrt}

% for BibTeX - sorted numerical labels by order of

% first citation.

% A useful Journal macro

\def\Journal#1#2#3#4{{#1} {\bf #2}, #3 (#4)}

% Some useful journal names

\def\NCA{\em Nuovo Cimento}

\def\NIM{\em Nucl. Instrum. Methods}

\def\NIMA{{\em Nucl. Instrum. Methods} A}

\def\NPB{{\em Nucl. Phys.} B}

\def\PLB{{\em Phys. Lett.}  B}

\def\PRL{\em Phys. Rev. Lett.}

\def\PRD{{\em Phys. Rev.} D}

\def\ZPC{{\em Z. Phys.} C}

% Some other macros used in the sample text

\def\st{\scriptstyle}

\def\sst{\scriptscriptstyle}

\def\mco{\multicolumn}

\def\epp{\epsilon^{\prime}}

\def\vep{\varepsilon}

\def\ra{\rightarrow}

\def\ppg{\pi^+\pi^-\gamma}

\def\vp{{\bf p}}

\def\ko{K^0}

\def\kb{\bar{K^0}}

\def\al{\alpha}

\def\ab{\bar{\alpha}}

\def\be{\begin{equation}}

\def\ee{\end{equation}}

\def\bea{\begin{eqnarray}}

\def\eea{\end{eqnarray}}

\def\CPbar{\hbox{{\rm CP}\hskip-1.80em{/}}}

%temp replacement due to no font

%%%%%%%%%%%%%%%%%%%%%%%%%%%%%%%%%%%%%%%%%%%%%%%%%%

%                                                %

%    BEGINNING OF TEXT                           %

%                                                %

%%%%%%%%%%%%%%%%%%%%%%%%%%%%%%%%%%%%%%%%%%%%%%%%%%

%\begin{document}

\vspace*{4cm}

\title{MAXIMUM ORBITAL FREQUENCIES FOR ROTATING QUARK STARS AND 
NEUTRON STARS---IMPLICATIONS FOR  THE  RESONANCE THEORY OF kHz QPOs}

\author{D. GONDEK-ROSI\'NSKA$^{1,2}$ and  W. KLU\'ZNIAK$^{3,4}$ }
\address{$^1$ LUTH, FRE 2462 du C.N.R.S., Observatoire de Paris,
 F-92195 Meudon Cedex, France,\\
$^2$ Nicolaus Copernicus Astronomical Center, Bartycka 18,
 00-716 Warszawa, Poland,\\
 $^3$ Institute of Astronomy, University of Zielona G\'ora, 
ul. Lubuska 2, Zielona G\'ora, Poland\\
$^4$ Institut d'Astrophysique de Paris, 98bis Boulevard Arago, 75014 Paris
France
}

\maketitle\abstracts{The spin-up of a neutron star crucially
depends on the maximum orbital frequency around it, as do a host
of other high energy accretion phenomena in low mass X-ray
binaries, including quasi-periodic oscillations (QPOs) in the
X-ray flux.  We compare the maximum orbital frequencies  for
MIT-bag quark stars and for neutron stars modeled with the FPS
equation of state. The results are based on relativistic 
calculations of constant baryon sequences of uniformly rotating
strange star models, and are presented as a function of the
stellar rotational frequency. The marginally stable orbit is
present outside quark stars for a wide range of parameters, but
outside the FPS neutron stars it is present only for the highest
values of mass. This allows a discrimination between quark stars
and neutron stars in the resonance theory of kHz QPOs.} 

\section{Introduction}

If the recently discovered kHz QPOs~\cite{Klis00}
 are a manifestation of strong gravity, they may be
used to constrain the external metric of the compact source and the
equation of state of matter at supranuclear densities.  Klu{\'z}niak
et al.\cite{KluznMW90}  
suggested that the mass of a neutron star may be derived
by observing the maximum orbital frequency, if it occurs in the
marginally stable orbit, and that the low frequency QPOs occurring in
X-ray pulsars will have their counterpart in LMXBs, at frequencies in
the kHz range. Such kHz QPOs have indeed been discovered and their
frequency used to derive mass values of about $2M_\odot$, under the
stated assumption.\cite{Kaaret97,Zhang98,Kluzn98,Bulik00}
 In fact, the QPO frequency may correspond to
a larger orbit, and hence a smaller mass.\cite{KluznA02}

The question whether quark stars may have maximum orbital frequencies
as low as the observed kHz QPO frequencies has also been investigated.
Bulik et al.\cite{BGK99a,BGK99b} showed that slowly rotating strange
stars, described by the simple MIT bag model with massless and
non-interacting quarks, have orbital frequencies at the marginally
stable orbit higher than the maximum frequency of 1.07  kHz in
4U 1820-30, reported by Zhang et al.\cite{Zhang98}.  However, the ISCO
frequencies can be as low as 1 kHz  
when more sophisticated models of quark matter
(with massive strange quarks and lowest order QCD interactions)
 and/or rapid stellar rotation are taken into
 account.\cite{StergKB99,GondeSBKG01,ZduniBKHG00,ZdunHGG00}
The  lowest orbital frequency at the ISCO was found
to be attained either for non-rotating massive configurations 
close to their  maximum mass limit, or for
configurations at the equatorial mass-shedding limit
for a broad range of stellar masses.
The effect of the crust, if present,
 has been investigated for normal evolutionary
sequences.\cite{ZdunHG01}  Typically the
crust increases the maximum orbital frequency at the Keplerian
limit.

If the strange star configurations described by Dey et. al.\cite{Dey98}
are allowed, the rotational and maximal orbital frequencies
are much higher~\cite{Datta00,GondeBZGRDD00,GondeBKZG01} 
than those for neutron star models, or for  the MIT-bag models of quark stars. 
The maximum orbital frequencies for the Dey et al. models
are always higher than the kHz QPO frequencies observed to date,
 and higher than 1.5 kHz for  stars with masses greater than $1 \, M_\odot$.

%%%%%%%%%%%%%%%%%%%%%%%%%%%%%%%%%%%%%%%%%%%%%%%%%%%%%%%%

\begin{figure}
%\rule{5cm}{0.2mm}\hfill\rule{5cm}{0.2mm}
\vskip 2.5cm
%\rule{5cm}{0.2mm}\hfill\rule{5cm}{0.2mm}
\psfig{figure=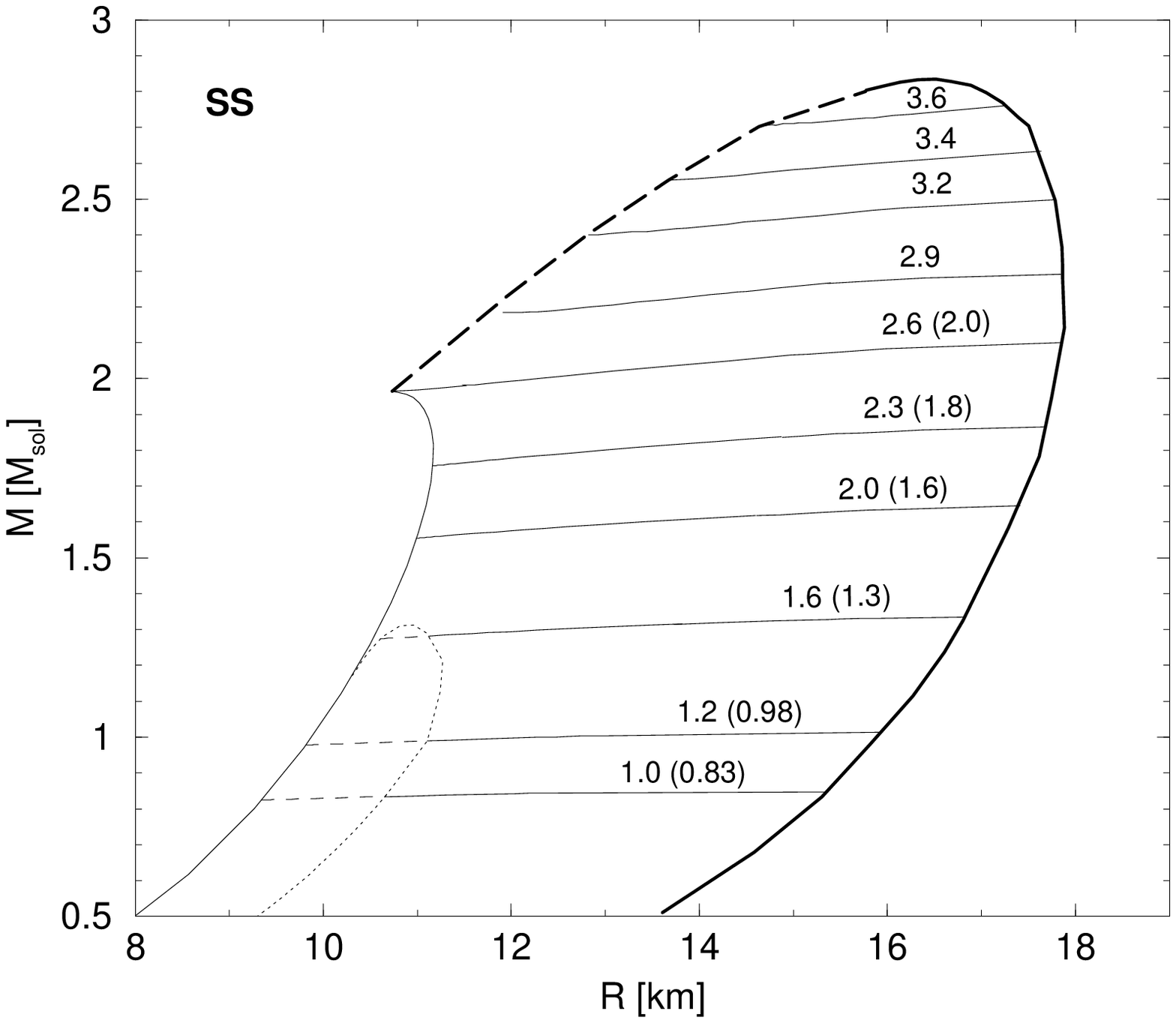,height=2.5in}\qquad
\psfig{figure=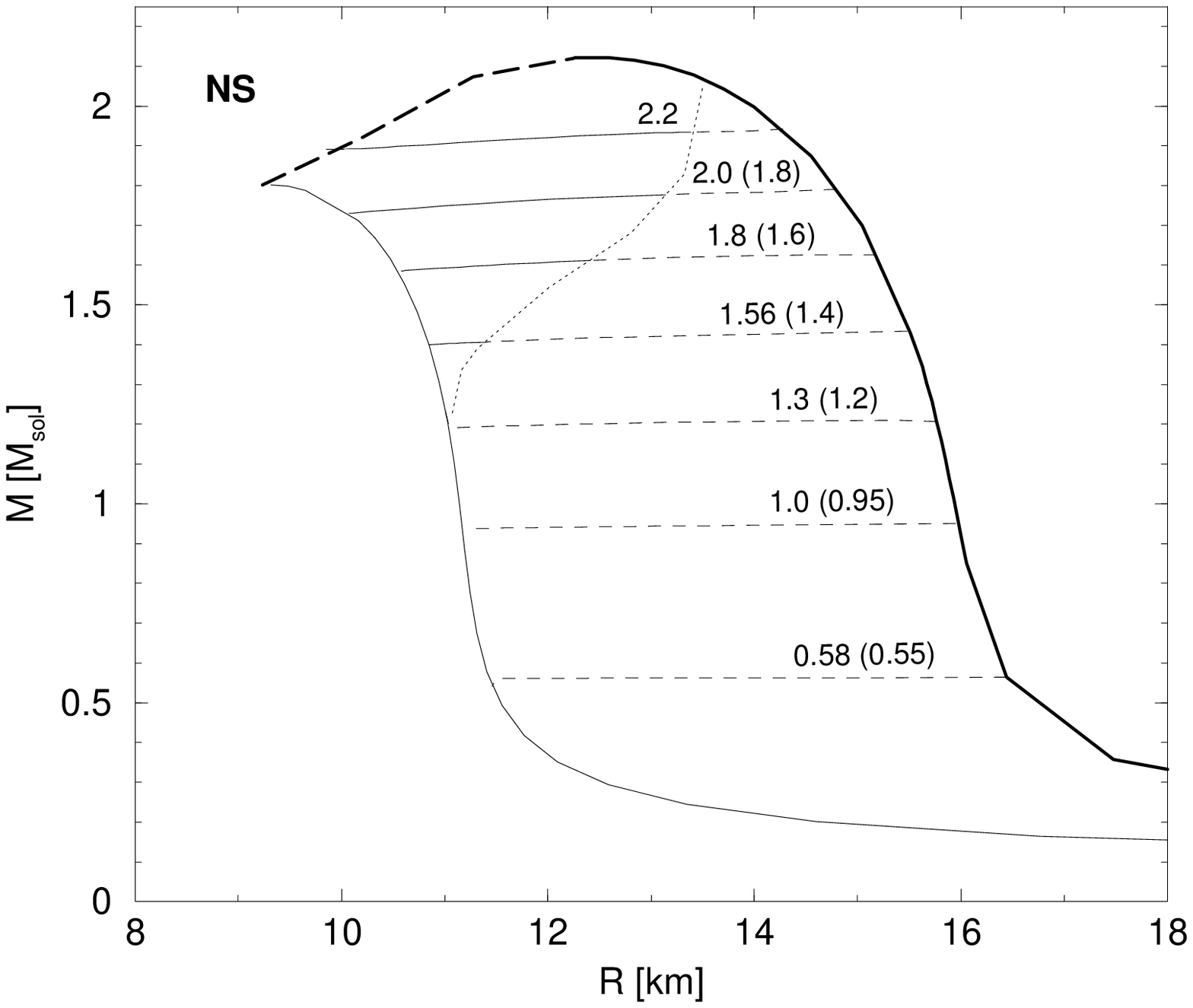,height=2.45in}
\caption{Gravitational mass versus radius for sequences of constant
baryon mass.  Each sequence is labeled by this baryon mass in solar
mass units, as well as (in parentheses) by the gravitational mass of
the static configuration, if it exists. The angular momentum increases
along each sequence from $J=0$ for static configurations, or $J_ {\rm
min}$ for supramassive stars, to $J_{\rm max}$ for the mass-shedding
limit (thick solid line).  The thick dashed lines indicate stars
marginally stable to axisymmetric perturbations.  The left panel shows
quark stars modeled with the MIT-bag e.o.s. for massless
non-interacting quarks. The right panel: neutron stars modeled with
the FPS e.o.s. The short-dashed lines correspond to rotating models
for which the marginally stable orbit does not exist. The dotted lines
separate the models with and without a marginally stable orbit.}
\label{fig:MR}
\end{figure}
%%%%%%%%%%%%%%%%%%%%%%%%%%%%%%%%%%%%%%%%%%%%%%%%%%%%%%%%%%%%%%

\begin{figure}
%\rule{5cm}{0.2mm}\hfill\rule{5cm}{0.2mm}
\vskip 2.5cm
%\rule{5cm}{0.2mm}\hfill\rule{5cm}{0.2mm}
\psfig{figure=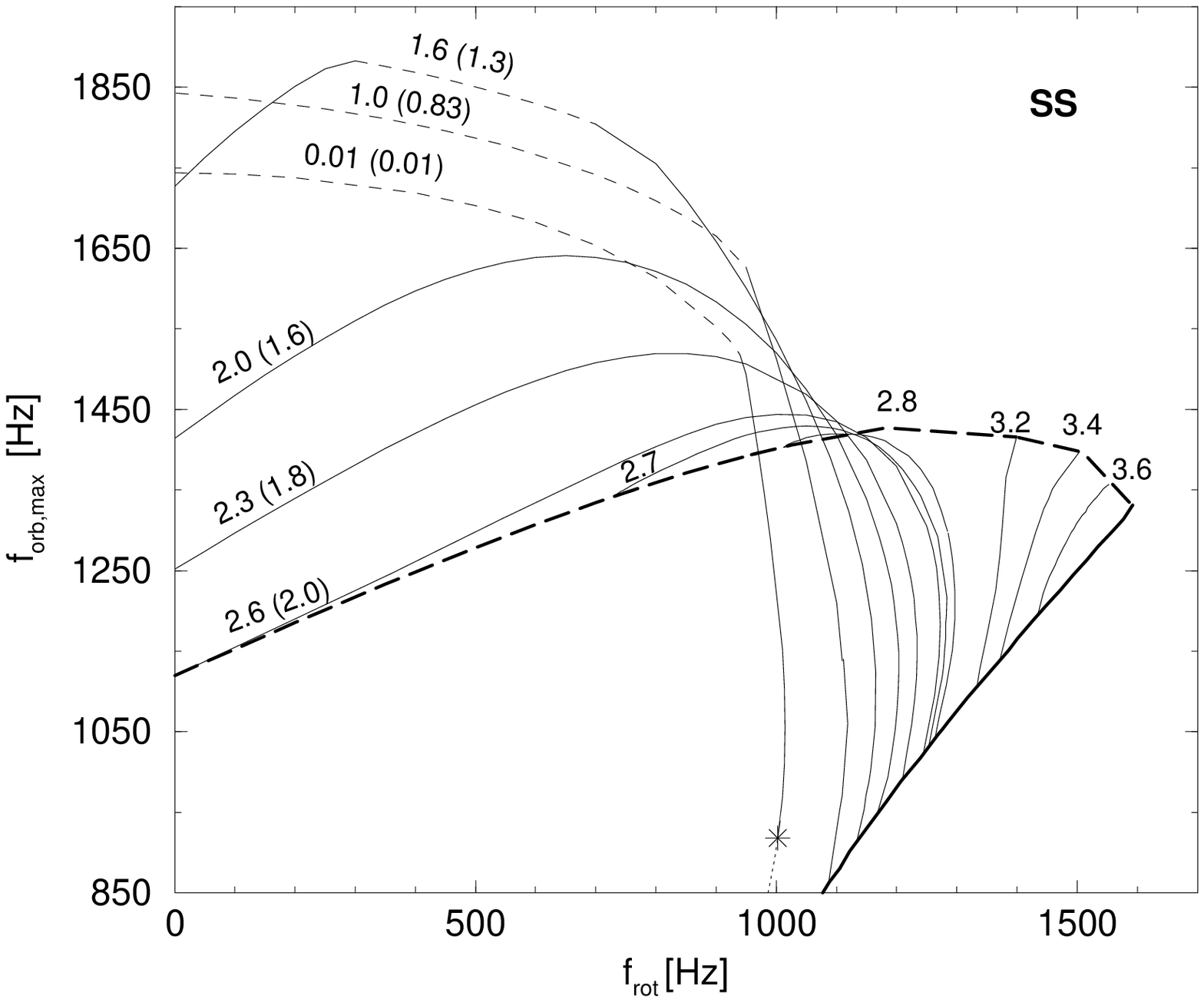,height=2.5in}\qquad
\psfig{figure=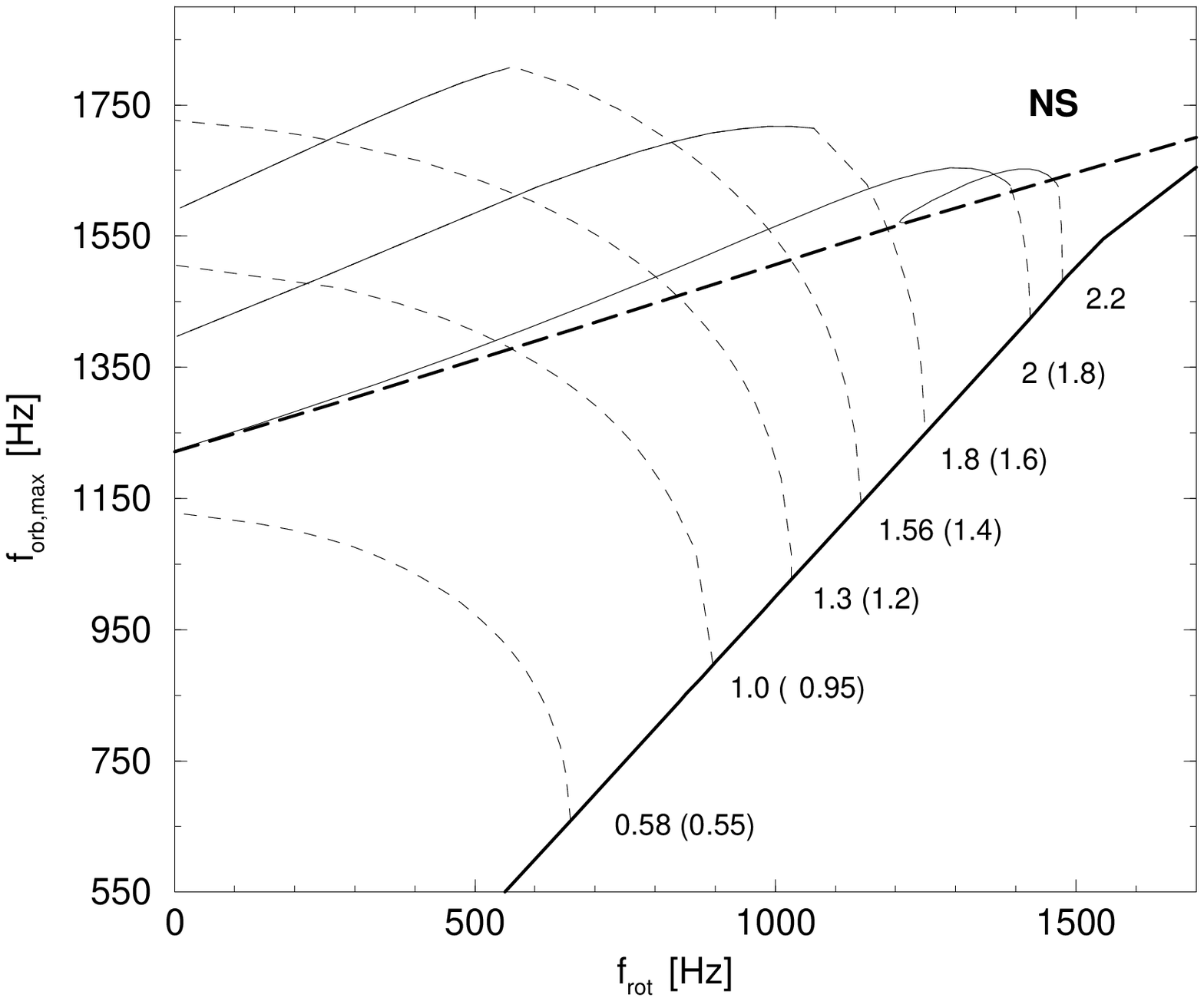,height=2.5in}
\caption{Maximum orbital frequency vs. the frequency of rotation
$\Omega/2\pi$ for the sequences of Fig.~1.  One sequence for a very
low mass quark star (with $M_{\rm b}=0.01 M_\odot$) is shown, the
critical point on this sequence for Newtonian dynamical instability to
non-axisymmetric perturbations is indicated by an asterisk and
dynamically unstable configurations are denoted with the dotted line.
The various dashed lines have the same meaning as in Fig.~1}
\label{fig:fmsf}
\end{figure}

\section{Results}

Here we summarize our studies of maximal orbital frequency of rotating
quark stars. We focus on neutron stars modeled with the FPS
e.o.s.\cite{Lorenz93}, and on quark stars described
by the simplified MIT bag model ($\alpha_{\rm c}=m_s=0$ and 
$B=60\,{\rm MeV/fm^3}$). Similar results are obtained for other models of
quark stars~\cite{GondeBKZG01,GondeSBKG01,ZdunHGG00,Z00} 
and neutron stars~\cite{CST94,Miller98}.   We have
performed the calculations using two different highly accurate
relativistic codes.\cite{GourgHLPBLM99,StergF95}

A comparison between the relevant properties of rotating quark stars
and neutron stars described by the FPS equation of state is afforded
by Figures 1 and 2. The dotted lines separate the models with and
without a co-rotating marginally stable orbit.

The orbits around neutron stars~\cite{CST94,Miller98} have different
properties than those around quark stars (Figures 1, 2).  We see that
for quark stars with moderate and high baryon masses the marginally
orbit is always present at any rotation rate.  For intermediate mass
quark stars, stable orbits extend down to the stellar surface for
moderate rotation rates (the short-dashed models), but the gap is
present for either slowly or rapidly rotating stars.  For the lowest
mass stars, the relativistic gap~\cite{KluznW85} is present only at
high rotation rates.

Note that the marginally stable orbit is present also for rapidly
rotating quark stars of very low mass - it has recently been
discovered that the innermost stable circular orbit exists also in
Newtonian gravity, the gap between the marginally stable orbit and the
stellar surface being produced by the oblateness of the rapidly
rotating low-mass quark star.\cite{AmsteBGK02,KluznBG00,ZdunG01} This
effect of oblateness seems to be responsible for the ``pushing
outwards'' of the marginally stable orbit even for massive rotating
quark stars (compare the discussion in Stergioulas et
al.\cite{StergKB99}).

Another consequence is that the period and the mass of rotating quark
stars cannot be even approximately inferred from the orbital
frequencies alone---the same frequency in the innermost stable orbit
(e.g., 1.25 kHz for the model presented) is obtained for quark stars
with rotational periods ranging from infinity to about 0.6 ms, and the
mass ranging from that of a planetoid to about three solar
masses.\cite{KluznBG00,GondeSBKG01}  One should note however
that quite likely the most rapidly rotating stars are
unstable.\cite{GondeSBKG01,GondeGH02}

\section{Astrophysical applications}

To illustrate how these results may be used to constrain the e.o.s. of
dense matter, let us assume that the results of Figs.~1 and 2 are
representative of quark stars and neutron stars.  The recently
proposed theory of resonant origin of kHz QPOs~\cite{KluzA01,AbramK01}
received strong observational support with the discovery of two black
hole sources in which the observed QPO frequencies are in a 2:3
ratio.\cite{Remil02} According to that theory,\cite{KluznA02} the high
frequency QPOs which come in pairs arise in an accretion disk,
presumed to be geometrically thin, as a result of parametric resonance
between the radial and vertical epicyclic frequencies. For a given
metric, this occurs at a specific radius.  In the Schwarzschild
metric, this resonance occurs at the radius $r_{2:3}=16.2\,{\rm
km}\times M/M_\odot$ and for moderately rotating neutron stars at
somewhat smaller radii.

For illustrative purposes suppose that in a star known to have a
 rotational period of no less than a few milliseconds, a QPO at 1.37
 kHz is detected and identified with the vertical epicyclic frequency
 at the resonant radius $r_{2:3}$. It follows, that the mass of the
 star must be $M<0.67M_\odot$ and the radius of the star must satisfy
 $R<r_{2:3}<10.9\,{\rm km}$.  The Schwarzchild values would be
 $M=0.6M/M_\odot$ and $R<9.7\,{\rm km}$, respectively.  As is evident
 from the figures, these constraints are not satisfied for any of the
 neutron star models (for the FPS equation of state), but are easily
 satisfied by the quark star models.  We conclude that, at least in
 principle, one can use the difference in the orbital properties of
 quark and neutron stars to distinguish observationally between the
 two classes of objects.

\section*{Acknowledgments}

This work has been funded by the following grants: KBN grants
 5P03D01721 and 2P03D02117; the Greek-Polish Joint Research and
 Technology Program EPAN-M.43/2013555 and the EU Program ``Improving
 the Human Research Potential and the Socio-Economic Knowledge Base''
 (Research Training Network Contract HPRN-CT-2000-00137).

\end{document}